\documentstyle[12pt]{article}
\def\be{\begin{equation}}
\def\ee{\end{equation}}

\def\ba{\begin{array}{c}}
\def\ea{\end{array}}
\def\p{\partial}
\begin{document}

\titlepage
\vspace*{2cm}

 \begin{center}{\Large \bf
  Relativistic  supersymmetric quantum mechanics based on
 Klein-Gordon equation
   }\end{center}

\vspace{10mm}

 \begin{center}
Miloslav Znojil

 \vspace{3mm}

\'{U}stav jadern\'e fyziky AV \v{C}R, 250 68 \v{R}e\v{z}, Czech
Republic\footnote{e-mail: znojil@ujf.cas.cz}

\end{center}

\vspace{5mm}


\section*{Abstract}

Witten's non-relativistic formalism of supersymmetric quantum
mechanics was based on a factorization and partnership between
Schr\"{o}dinger equations.  We show how it accommodates a
transition to the partnership between relativistic Klein-Gordon
equations.

\vspace{5mm}

PACS   03.65.Fd; 03.65.Ca; 03.65.Ge; 11.30.Pb; 12.90.Jv

\newpage

\section{Introduction}

The search for parallels between non-relativistic Schr\"{o}dinger
equation
 \be
  i\p _t \psi^{(SE)}(x,t) = \hat{h}^{(SE)}\,\psi^{(SE)}(x,t)\,
 \label{SE}
 \ee
and Klein-Gordon equation referring to relativistic kinematics in
zero-spin case,
 \be
 \
 \left( i\,\p_t\right )^2 \Psi^{(KG)}(x,t)
 = \hat{H}^{(KG)}\,\Psi^{(KG)}(x,t)\,
 \label{jednar}
 \ee
has met one of its important successes in the work by Feshbach and
Villars \cite{FV} who found that a close connection between eqs.
(\ref{SE}) and (\ref{jednar}) was mediated by the degree-lowering
method of Peano and Baker \cite{Ince}. In such an approach, one
treats the first derivative $\varphi_1^{(PB)}(x,t) =
i\,\p_t\,\Psi^{(KG)}(x,t)$ of a wavefunction as an independent
quantity complemented by the wavefunction itself,
$\varphi_2^{(PB)}(x,t)=\Psi^{(KG)}(x,t)$. In this notation,
Klein-Gordon equation (\ref{jednar}) may be re-written in
Schr\"{o}dinger-like form
 \be
 i\,
 \p_t
 \left (
 \ba
 \varphi_1^{(PB)}(x,t)\\ \varphi_2^{(PB)}(x,t)
 \ea
 \right )
=
 \hat{h}^{(PB)}\,
 \cdot \left (
 \ba
 \varphi_1^{(PB)}(x,t)\\ \varphi_2^{(PB)}(x,t)
 \ea
 \right ), \ \ \ \ \ \ \ \
 \hat{h}^{(PB)}=\left (
 \begin{array}{cc}
 0&\hat{H}^{(KG)}\\
 1&0
 \ea
 \right )
 \,.
 \label{jdvar}
 \ee
The manifest non-Hermiticity of the relativistic Peano-Baker
Hamiltonian $\hat{h}^{(PB)}$ in the latter equation seems to
obstruct its compatibility with quantum mechanics. A new hope has
been provided by the pioneering letter by Bender and Boettcher
\cite{BB} who studied some non-Hermitian non-relativistic
Hamiltonians $\hat{h}^{(SE)}$ and emphasized that they may often
possess a real, ``observable" spectrum \cite{citea}. This
immediately caused a perceivable intensification of interest in
all the models with real spectra which satisfy a weakened (or
``pseudo-") Hermiticity condition \cite{Dirac},
 \be
 \left [\hat{h}^{(SE)}\right ]^\dagger =
 {G}\,\hat{h}^{(SE)}\,{G}^{-1},
 \ \ \ \ \ \ {G} = {G}^\dagger\,.
 \label{etaf}
 \ee
The Bender's and Boettcher's reduction of the Hermitian
conjugation to the mere action of the time-reversal operator
${\cal T}$ proved as productive as their simplifying
identification of ``the metric" $G$ with the parity operator
${\cal P}$. More or less systematic description of many new ${\cal
PT}$ symmetric models followed, and an unexpectedly robust reality
of their spectra has been confirmed \cite{cite}. Still, within an
emerging ${\cal PT}$ symmetric and/or pseudo-Hermitian quantum
mechanics, the thorough and lasting attention had to be paid to
the unpleasant, indefinite character of the conserved pseudo-norm
\cite{cons}.

The progress was quick. Firstly, within a broad subset of the
exactly solvable ${\cal PT}$ symmetric non-relativistic models the
existence of a specific symmetry (called quasi-parity ${\cal Q}$
\cite{ptho}) has been revealed, with the main capability of an
elimination of the indeterminacy of the sign in the pseudo-norm.
Temporarily, the existence of the quasi-parity seemed restricted
to the exactly solvable models only \cite{Quesne}. Fortunately, in
2002, the ultimate resolution of the puzzle was described by A.
Mostafazadeh \cite{AM} and, independently, by Bender et al
\cite{BBJ}. On an abstract algebraic level, the former author
imagined that an elementary clarification of the situation may be
based on the fact (well known, e.g., in nuclear physics
\cite{Geyer}) that each diagonalizable operator $\hat{h}^{(SE)}$
with the real spectrum may be assigned {\em many } alternative
metrics ${G}_j$. Some of them {\em must} be positive definite (let
us denote them ${G}_+$) in many cases of immediate interest, with
a particular exemplification $G_+ = {\cal QP}$ using quasi-parity
in the solvable examples as mentioned above. In parallel, Bender
et al \cite{BBJ} introduced the very similar metric $G_+ = {\cal
CP}$ where, in contrast to the previous case, no solvability
assumption was needed. Their additional symmetry ${\cal C}$
carries the name of a ``charge" of the system and is defined by
its spectral representation. The practical feasibility of the
perturbative and/or variational constructions of ${\cal C}$ has
been demonstrated in ref. \cite{Wang}.

As long as the relativistic Klein-Gordon eq.~(\ref{jdvar}) is
characterized by the two-by-two form of the pseudo-Hermiticity of
its Hamiltonian,
 \be
 \left [\hat{h}^{(PB)} \right ]^\dagger
  = {\cal G}\,\hat{h}^{(PB)}\,{\cal G}^{-1}, \ \ \ \ \ \ \ \
 {\cal G}=\left (
 \begin{array}{cc}
 0&{G}\\
 {G}&0
 \ea
 \right )
 \,
\label{neheri}
 \ee
we {\em know} that it also possesses an alternative positive
metric ${\cal G}_+$ which induces its {\em fully consistent}
probabilistic interpretation. As long as such an argument has a
mere {\em implicit} character, A. Mostafazadeh offered, in his
three very recent papers \cite{p57}, also an {\em explicit}
example of a sufficiently simple ``physical" metric ${\cal G}_+$.
Choosing just the most elementary free Klein-Gordon operator
$\hat{H}^{(KG)}_0=-\p^2_x+m_0^2>0$ he described thoroughly the
properties of the resulting inner product which is conserved and
{\em positive-definite} on the whole space of the solutions of eq.
(\ref{jdvar}). This means that the free Klein-Gordon equation is
capable to guarantee the existence of the conserved norm, etc. A
path towards a satisfactory and consistent relativistic quantum
mechanics is open. In this sense we feel encouraged to return to
eq.~(\ref{jdvar}), say, with a nontrivial model interaction based
on a ``minimal" replacement of the constant $m_0^2$ by its
coordinate-dependent, ``scalar" potential or ``local
effective-mass" generalization $m^2(x) = m_0^2+V(x)$.

In what follows we intended to develop a most natural Klein-Gordon
generalization of the Witten's supersymmetric quantum mechanics
(SUSY QM, \cite{Witten,CKS}). For the time being let us choose
just the trivial partitions $G = I$ in eq. (\ref{neheri}) (i.e.,
Hermitian sub-Hamiltonians $\hat{H}^{(KG)}$) and summarize our
forthcoming effort as a construction which circumvents the three
most serious obstacles which might mar a naive SUSY construction
in Klein-Gordon case.

\begin{itemize}

\item
(a) the spectrum may become ``too rich" for supersymmetrization
purposes;

\item
(b) the Klein-Gordon supersymmetric  construction will be
nontrivial because, in contrast to the traditional
non-relativistic case, one of the partner equations exhibits an
anomalous, ``unavoided" crossing of its energy levels;

\item (c) the Klein-Gordon supersymmetrization  will require a
new, not yet known type of a factorization of the operators.

\end{itemize}

 \noindent
In the sequel, the point (a) will be dealt with in section
\ref{dveje}. We shall see there that many ``too strongly
interacting" operators $\hat{h}^{(PB)}$ may possess complex and,
hence, physically meaningless eigenvalues. This obstacle is,
fortunately, trivial since in such a regime the forces in action
must be very strong so that one encounters a phenomenologically
motivated limits of validity of quantum mechanics. One is forced
to use many more degrees of freedom which may be made available,
most typically, in quantum field theory.

The study of point (b) will be initiated in section \ref{triva}
where we recollect the main steps of supersymmetrization in
non-relativistic context and remind the reader that the
supersymmetry requires that the ground-state energy vanishes. In
the relativistic Klein-Gordon setting where the first unavoided
crossing of quantum levels occurs precisely at the vanishing
energy, this requirement becomes highly nontrivial.

The core of our present message concerns the point (c) and will be
discussed in section \ref{valgebra}. We shall propose a new type
of a factorization of our Hamiltonians, the feasibility of which
will be achieved via an additional partitioning of our Hilbert
space(s). We shall emphasize the intimate relationship of the
properties of our new factorization to the above-mentioned
existence of the level crossings, connected also to the numerical
concept of the Jordan blocks in non-Hermitian matrices
\cite{Iochvidov}.

The summary of our construction will be given in section
\ref{fiveth}.

\section{Non-Hermiticity and spectra of Klein-Gordon equations
\label{dveje} }

\subsection{Auxiliary partitioning of Hilbert space
}

Let us assume that the operator $\hat{h}^{(PB)}\neq \left (
\hat{h}^{(PB)} \right )^\dagger$ in the Peano-Baker form
(\ref{jdvar}) of our Klein-Gordon equation (\ref{jednar}) does not
depend on time and remains ``regular", $\hat{h}^{(PB)}
=\hat{h}^{(PB)}_{(R)}$ having no pathological complex or vanishing
eigenvalues. Then, the standard Fourier-type ansatz for
wavefunctions,
 \be
 \varphi_1^{(PB)}(x,t) = \int U(x,E)\,e^{-i\,E\,t}\,d \mu(E), \ \ \ \
 \varphi_2^{(PB)}(x,t) = \int D(x,E)\,e^{-i\,E\,t}\,d \mu(E)\,
 \ee
reduces eq. (\ref{jednar}) to its linear algebraic eigenvalue
version
 \be
 \left (
 \begin{array}{cc}
 0&\hat{H}^{(KG)}\\
 I&0
 \ea
 \right )
 \cdot \left (
 \ba
 U(x,E)\\ D(x,E)
 \ea
 \right )=
 E\,
 \left (
 \ba
 U(x,E)\\ D(x,E)
 \ea
 \right )
 \,.
 \label{jepiodnadfr}
 \ee
The solutions have to be sought in the relativistic
Feshbach-Villars Hilbert space composed of two equivalent
subspaces, ${\cal R}_{(R)}={\cal H}_{(R)}\oplus {\cal H}_{(R)}$.
Their components $U(x,E)$ (``up") and $D(x,E)$ (``down") belong to
the identical subspaces ${\cal H}_{(R)}$ with elements which will
be denoted by the curly kets $| \cdot \}$. In this notation, the
second row of the partitioned equation (\ref{jepiodnadfr}) implies
that $ |U\} =E\, |D\}$. Formally, this solves eq.
(\ref{jepiodnadfr}) by its reduction to the equation
 \be
 \hat{H}^{(KG)}\,|
 D_n^{(\pm)}\}=\varepsilon_n\, |D_n^{(\pm)}\},
 \ \ \ \ \ \ \ \ n = 0, 1, \ldots
 \,
 \label{upeddfr}
 \ee
in ${\cal H}_{(R)}$. Under our above assumption that
$\varepsilon_n> 0$, the original energies form pairs $E_n^{(\pm)}
= \pm \sqrt{\varepsilon_n}$ and remain real. In many cases, the
set of kets $|D_n^{(+)}\}\}$ (equal, up to a multiplicative
factor, to $|D_n^{(-)}\}\}$) will be complete in ${\cal H}_{(R)}$.
In contrast, the ``up" components $|U_n^{(\pm)}\} = E_n^{(\pm)}\,
|D_n^{(\pm)}\}$ depend on the superscript and define the {\em
pairs} of independent eigenvectors of eq. (\ref{jepiodnadfr})
numbered by the sign superscript {\em and} integers~$n$,
 \be
 \hat{h}^{(PB)}\,
 |n^{(\pm)} \rangle =
 E_n^{(\pm)}\,
 |n^{(\pm)} \rangle, \ \ \ \ \ \ \ \
 |n^{(\pm)} \rangle =
 \left (
 \ba
 \pm \sqrt{\varepsilon_n}\, |D_n^{(\pm)}\}\\ \, |D_n^{(\pm)}\}
 \ea
 \right )
 \,.
 \label{generated}
 \ee
We shall often assume that these kets span the whole relativistic
Hilbert space ${\cal R}$ (see Appendix A giving more details).

\subsection{An exactly solvable illustrative example
 \label{examplesc}}

For a class of the model Klein-Gordon operators
 \be
 H^{(KG)} = -\p^2_x + m^2(x), \ \ \ \ \ \ \ \ x \in I\!\!R
 \ee
where $m_0=m(\pm \infty)$ is the free asymptotic mass and
$c=\hbar=1$, the key assumption $\varepsilon_n> 0$ of our
preceding paragraph is easily satisfied for a sufficiently smooth
and small variations of the ``effective mass" $ m^2(x)$ at the
finite coordinates $x \in I\!\!R$. Such a choice of the model
enables us to discuss some of the consequences of making this
coordinate-dependence less and less smooth and/or bounded. For
this purpose let us pick up one of the most elementary models with
attraction,
 \be
 m^2(x)=m^2_0+\frac{B^2-A^2-A\omega}{\cosh^2\omega x}
 +\frac{B(2A+\omega) \sinh\,\omega x}{\cosh^2\omega x}
 \label{11}
 \ee
which is exactly solvable and shape-invariant \cite{CKS}. Its
exact solvability means that its bound-state solutions generated
by our Klein-Gordon equation (\ref{jepiodnadfr}) exist at (a
finite set of) the energies
 \be
 E_n^{(\pm)} = \pm \sqrt{m^2_0-(A-n\omega)^2}
 , \ \ \ \ \ \ n = 0, 1, \ldots, n_{max},
 \ \ \ \ n_{max}=entier[A/\omega]\,.
 \label{spektrsc}
 \ee
We may distinguish between the three regimes. In the first one
with $0 < A < m_0$, the strength of the force is weak and we have
the well-behaved eigenvalues at all the admissible indices $n$. In
the second, strong-coupling regime with $0 < m_0 < A$, a few
low-lying states suffer a collapse and acquire, formally, complex
energies, i.e., $Im(E_0^{(\pm)}) \neq 0$, $Im(E_1^{(\pm)}) \neq 0$
etc. We have certainly left the domain of quantum mechanics.

At the boundary point we have $ A= m_0>0$. During a limiting
transition $\varepsilon_0\to 0^+$ the two smallest eigenvalues
$E_0^{(\pm)}$ do merge and vanish. Under the name of an
exceptional point (EP, \cite{Heiss}) the energy $E_0=0$ still
represents a valid bound state because after one abbreviates
  \be
 y = y(x)=\sinh \omega x, \ \ \ s = A/\omega > 0, \ \ \ t =
 B/\omega\,,
 \ee
the wavefunctions pertaining to the set of the energies
$E_n^{(\pm)}$ may be {\em all} written in closed form in terms of
Jacobi polynomials,
 \be
 \langle x | D_n\}=i^n[1+y(x)^2]^{-s/2} \exp [ -t\,\arctan y(x)]
  \,P_n^{(it-s-1/2,-it-s-1/2)} [iy(x)] \,.
  \ee
In spite of appearances, all of them remain normalizable at all
$s>0$ so that also the solution with $E_0^{(\pm)}=\varepsilon_0=
0$ (in fact, just a single state) does not represent any
exception. All the wavefunctions appear to be purely real and
after we insert the abbreviation $y(x)=\sinh \omega x$ we get
 \be
  \langle x | D_0\}
  =(1+\sinh^2 \omega x)^{-s/2} \exp ( -t\,\arctan \sinh \omega x)
  , \ \ \
  \ee
 \be
  \langle x | D_1\}
  =(1+\sinh^2 \omega x)^{-s/2} \exp ( -t\,\arctan \sinh \omega x)
  \,
 \left (
  \frac{2s-1}{2}\,\sinh \omega x - t
  \right )
  , \ \ \ \ldots \,.
  \ee
One should be fully aware that in contrast to the robust character
of the wavefunctions, the vanishing energy of the EP state at
$n=0$ becomes highly sensitive to a perturbation of the potential.
An arbitrarily small downward shift of $m^2(x)$ converts the
degenerate ground-state energy $E_0^{(\pm)} =0$ into a complex
conjugate pair of purely imaginary values, leading to an obvious
physical instability.

\subsection{Action of the non-Hermitian $\hat{h}^{(PB)}$
to the left}

Due to the non-Hermiticity (\ref{neheri}) of $\hat{h}^{(PB)}$, it
is convenient to complement eq. (\ref{jepiodnadfr}) by its
analogue where the Hermitian conjugation of the operator
$\hat{h}^{(PB)}$ is considered,
 \be
 \left (
 \begin{array}{cc}
 0&I\\
\left [\hat{H}^{(KG)} \right ]^\dagger&0
 \ea
 \right )
 \cdot \left (
 \ba
 L^*(x,E)\\ R^*(x,E)
 \ea
 \right )=
 E^*\,
 \left (
 \ba
 L^*(x,E)\\ R^*(x,E)
 \ea
 \right )
 \,.
 \label{dvaiodnadfr}
 \ee
In what follows, we shall prefer the more natural form of the same
equation where the operator $\hat{h}^{(PB)}$ is simply assumed to
act to the left, i.e., where the usual transposed
vector-times-matrix multiplication convention is used,
 \be
 \ba
 \left [
 L(x,E), \,R(x,E)
 \right ]
 \cdot
 \\
 \left . \right .
 \ea
 \left (
 \begin{array}{cc}
 0&\hat{H}^{(KG)}\\
 1&0
 \ea
 \right )
 \ba
  =\,
 E\,\cdot\,
 \left [
 L(x,E), \,R(x,E)
 \right ]\,,
 \\
 \left . \right .
 \ea
 \label{jednfradfpr}
 \ee
We shall employ the double-bra-like symbols $\{ \{ \cdot |$ for
both the ``left"  $L(x,E)$ and the ``right"  $R(x,E)$ functions in
the dual of ${\cal H}$.  {\it Mutatis mutandis} we arrive at the
auxiliary definition of $\{\{R_n^{(\pm)}| = E_n^{ (\pm)}\, \{ \{
L_n^{(\pm)}|$ and reduce eqs. (\ref{dvaiodnadfr}) and
(\ref{jednfradfpr}) to the standard eigenvalue problem $\left [
\hat{H}^{(KG)} \right ]^\dagger |L_n^{(\pm)}\}\}\, =\left (
\varepsilon_n \right )^*\, |L_n^{(\pm)}\}\}$, i.e., under our
present conventions,
 \be
 \{ \{ L_n^{(\pm)}|\,\hat{H}^{(KG)}
 =\varepsilon_n\, \{ \{ L_n^{(\pm)}|
 , \ \ \ \ \ \ n = 0, 1, \ldots \,.
 \label{upretseddfr}
 \ee
The new solutions $\{ \{ L_n^{(\pm)}|$ emerge at the same
eigenvalues $\varepsilon_n$ as above. This means that we may
denote all the doubly partitioned left row eigenvectors in the
(dual of the) larger Hilbert space ${\cal R}$ by the double-bra
symbol,
 \be
 \langle \langle n^{(\pm)} | =
 \left (\,
 \{ \{ L_n^{(\pm)}|,\,
\pm \sqrt{\varepsilon_n\,}\, \{ \{ L_n^{(\pm)}|\,
 \right )
 \,.
 \label{leftgenerated}
 \ee
Further merits of this compactified notation are summarized in
Appendix A.

 \section{Non-relativistic supersymmetric quantum mechanics
 \label{triva} }

Non-relativistic supersymmetric quantum mechanics (SUSY QM, cf.
its comprehensive review \cite{CKS}) may be defined as a formalism
where the operator of energy (i.e., an essentially self-adjoint
Hamiltonian $H=H^\dagger$ defined in a space ${\cal S}$) happens
to coincide with a generator of a graded Lie algebra. For example,
we may incorporate such a $H$ in the $sl(1|1)$ multiplication
table
 \be
 \{Q,Q\}=
 \{Q^\dagger,Q^\dagger \}=0,\ \ \ \ \
 \{Q,Q^\dagger\}= H,  \ \ \ \
 \left [ H ,Q \right ] = \left [ H, Q^{\dagger} \right ] =0\,
 \label{antiko}
 \ee
containing commutators as well as anticommutators. The other two
generators (so called ``supercharges" $Q$ and $Q^\dagger$) remain
non-Hermitian, i.e., unobservable. The potential appeal of such a
non-Lie model in physics has been revealed, i.a., by E. Witten
\cite{Witten} who claimed that the analyses of similar systems
(tractable also as field theories in zero dimensions in particle
physics) could clarify the puzzling experimental absence of
``supermultiplets" containing both bosons {\em and} fermions. For
this reason, it sounds like a paradox that while the methodical
impact of the oversimplified SUSY QM on experimental {physics}
remained virtually negligible, the supersymmetry  using
superalgebra $sl(1|1)$ offers a lasting inspiration of theoretical
developments in quantum mechanics \cite{bylvalladolid}.

The representations of the superalgebra $sl(1|1)$ were extended to
non-Hermitian Hamiltonians \cite{Cann}. A particularly promising
continuation of this direction of research might eventually be
related to the relativistic extension of SUSY QM based on the use
of Klein-Gordon equations \cite{Exper}. Indeed, the more
traditional Hermitian constructions seem both too formal and too
narrow. Too formal because almost the whole story of the
traditional SUSY QM may be re-told as an application of
Schr\"{o}dinger's factorization method \cite{Stalhofen}, and too
narrow because even the simplest relativistic spin-one-half
Dirac-equation examples (cf. their sketchy review in section
Nr.~11 of ref. \cite{CKS}) do not seem to fit in the general
scheme. In what follows we shall try to address and weaken both of
these objections.

\subsection{Representations of superalgebra $sl(1|1)$
\label{Khare}}

The validity of the mixed commutation and anticommutation
relations (\ref{antiko}) in SUSY QM is guaranteed by the
two-by-two matrix choice of all the generators in question,
 \be
 Q=\left (
 \begin{array}{cc}
 0&0\\
 \hat{b}&0
 \ea
 \right ), \ \ \ \
 Q^\dagger=\left (
 \begin{array}{cc}
 0&\hat{b}^\dagger\\
 0&0
 \ea
 \right ), \ \ \ \
 H=\left (
 \begin{array}{cc}
 \hat{h}_L&0\\
 0&\hat{h}_R
 \ea
 \right )\,.
 \label{rovni}
 \ee
The upper (= left) sub-Hamiltonian $\hat{h}_L$ and its ``right"
partner $\hat{h}_R$ are both acting in separate Hilbert spaces
${\cal H}_{(L,R)}$ such that
 $
{\cal S}^{}_{}=
 {\cal H}_{(L)}\oplus
 {\cal H}_{(R)}
 $.
In these subspaces, both the sub-Hamiltonians are essentially
self-adjoint and may be expected to possess a spectral
representation, i.e., formally,
 $
  \hat{h}_{L,R}
  =\sum_{n}\,|n_{(L,R)}\rangle
  \,\varepsilon_{(L,R)}^{(n)}\,\langle n_{(L,R)} |\,
  $.
In addition, both of them must be factorizable, $ \hat{h}_{L}=
\hat{b}^\dagger \hat{b}$ and $ \hat{h}_{R}=
\hat{b}\hat{b}^\dagger$. This is one of the most important
consequences of the SUSY postulate (\ref{antiko}). For the same
reason, the two sets of wavefunctions are interconnected by the
chain rules
 \be
 \left |  n_{(R)}\right \rangle=const\cdot
 \hat{b}\,
 \left | (n+1) _{(L)}\right \rangle, \ \ \ \ \ n = 0, 1,\ldots
 \ee
 \be
 \left |  (n+1)_{(L)}\right \rangle=const'\cdot
 \hat{b}^\dagger\,
 \left | n _{(R)}\right \rangle, \ \ \ \ \ n = 0, 1,\ldots\,
 \ee
and, in the regime of the so called unbroken supersymmetry we may
often deduce that the sub-spectra are, up to the ground state, the
same,
 \be
  0=\varepsilon_{(L)}^{(0)}<
  \varepsilon_{(R)}^{(0)}=
  \varepsilon_{(L)}^{(1)}<
  \varepsilon_{(R)}^{(1)}=
  \varepsilon_{(L)}^{(2)}<
  \ldots\,.
  \label{isos}
 \ee
Although the L- and R-bases may be different in principle, we may
put
 \be
  \hat{b}
  =\sum_{m}\,\left |m_{(R)}\right \rangle
  \cdot \beta_m\cdot \left \langle (m+1)_{(L)} \right |\,,
  \ \ \ \ \ \ \ \ \
  \hat{b}^\dagger
  =\sum_{p}\,\left |(p+1)_{(L)}\right \rangle
  \cdot\beta^*_p \cdot \left \langle p_{(R)} \right |\,
  \ee
and define easily also the basis and spectral representation of
our self-adjoint super-Hamiltonian $H$ in a larger Hilbert space
${\cal S}$.

In many applications, the basis is not employed and the
factorization recipe is specified in terms of the following linear
differential operators,
 \be
 \hat{b}=\p_x+W(x), \ \ \ \ \ \
 \hat{b}^\dagger=-\p_x+W(x), \ \ \ \ \ \
 \hat{h}_{L/R}=
 -\p^2_x+W^2(x) \mp W'(x)\,.
 \label{operators}
 \ee
Here, the so called superpotential $W(x)$ may be derived from an
arbitrary auxiliary input ground-state wavefunction,
 \be
 W(x)=-\frac{\psi_0'(x)}{\psi_0(x)}\,
 , \ \ \ \ \ \ \ {\psi_0(x)} \in
 {\cal
 H}_{(L)}
 \,.
 \ee
For illustration one may select the harmonic-oscillator
ground-state $\psi^{(HO)}_0(x) \sim \exp \left ( -x^2/2\right )$
giving the linear $W^{(HO)}(x) =x$ and the two partner
Hamiltonians $\hat{h}_{L/R}^{(HO)}= -\p^2_x+x^2 \mp 1$. Their pair
forms the desired super-Hamiltonian $H^{(HO)}$ with both the
bosonic {\em and} fermionic features.

\subsection{Example: Bosonic plus fermionic
 harmonic oscillator \label{exampleh}}

The simultaneous use of the language of bases and differential
operators is useful. In particular, our explicit knowledge of the
HO basis may help to clarify the essence of the factorization
property since in the infinite-dimensional matrix notation with
 \be
 |  0_{(L)}^{(HO)}\rangle=|  0_{(R)}^{(HO)}\rangle=
 \left (
 \ba
 1\\0\\0\\
 \vdots
 \ea
 \right ),\ \ \ \
 |  1_{(L)}^{(HO)}\rangle=|  1_{(R)}^{(HO)}\rangle=
 \left (
 \ba
 0\\1\\0\\
 \vdots
 \ea
 \right ), \ldots\,
 \label{trinact}
 \ee
we may simply put
 \be
 \hat{b} =
 \left (
 \begin{array}{ccccc}
 0&\beta_0&0&\ldots&\\
 0&0&\beta_1&0&\ldots\\
 0&0&0&\ddots&
 \ea
 \right ), \ \ \ \ \
 \hat{b}^\dagger =
 \left (
 \begin{array}{cccc}
 0&\ldots&&\\
 \beta_0^*&0&\ldots&\\
 0&\beta_1^*&0&\ldots\\
 0&0&\ddots&
 \ea
 \right )
 \label{examplea}
 \ee
and choose
 \be
 \beta_n=\sqrt{\varepsilon_{(L)}^{(n+1)}}=
  \sqrt{\varepsilon_{(R)}^{(n)}}\,.
  \ee
This facilitates the verification of the SUSY-based isospectrality
(\ref{isos}),
 \be
 \{ \varepsilon_{(L)}^{(n)} \} =
 \{E_L^{(HO)}\} = \{0, 2, 4, 6, \ldots\},
 \ \ \ \ \ \
 \{ \varepsilon_{(R)}^{(n)} \} =
 \{E_R^{(HO)}\} = \{ 2, 4, 6, \ldots\}\,.
 \ee
Also the physics ``hidden" besides the SUSY mathematics finds an
impressive Fock-space re-interpretation (as described in section
2.1 of review~\cite{CKS}) in the full Hilbert space ${\cal
S}^{(HO)}$ spanned by the two-index ket vectors
 \be
 |0,0\rangle=
 \left (
 \ba
 |0\rangle\\
 0
 \ea
 \right ), \ \ \ \
 |0,1\rangle=
 \left (
 \ba
 0\\
 |0\rangle
 \ea
 \right ), \ \ \ \
 |1,0\rangle=
 \left (
 \ba
 |1\rangle\\
 0
 \ea
 \right ), \ \ \ \
 |1,1\rangle=
 \left (
 \ba
 0\\
 |1\rangle
 \ea
 \right )
 \ee
etc, i.e., vectors $|n_b,m_f\rangle$ where the non-negative
integer $n_b = 0, 1, 2, \ldots$ may be understood as a number of
bosons in the system while the absence/presence of a fermion
(which obeys the Pauli exclusion principle) is characterized by
the second index with the mere two admissible values $n_f = 0 $
and $n_f =1$. In the other words, the creation or annihilation of
a boson is mediated by $\hat{b}^\dagger$ or $\hat{b}$,
respectively, while the operators of the creation and annihilation
of the fermion are just the elementary two-by-two matrices
 \be
 F^\dagger =
\left (
 \begin{array}{cc}
 0&0\\1&0
 \ea
 \right ), \ \ \ \ F=
 \left (
 \begin{array}{cc}
 0&1\\0&0
 \ea
 \right )
 \ee
which enter also the above definitions of the supercharges, $ Q =
F^\dagger\cdot \hat{b}$ while $ Q^\dagger = \hat{b}^\dagger\cdot F
$. Marginally, let us note that in the spirit of the simplified
notation of eq. (\ref{trinact}) we may also write
 \be
 |0,0\rangle=
 \left (
 \ba
  1\\
  0\\
  \vdots \\
  \cdot\cdot\cdot\\
 0\\
 \vdots
 \ea
 \right ), \ \ \ \
 |0,1\rangle=
 \left (
 \ba
  0\\0\\
  \vdots \\
\cdot\cdot\cdot\\
 1\\
 0\\
 \vdots
 \ea
 \right ), \ \ \ \
 |1,0\rangle=
 \left (
 \ba
 0\\
  1\\
  0\\
  \vdots \\
 \cdot\cdot\cdot\\
 0\\
 \vdots
 \ea
 \right ), \ \ \ \
 |1,1\rangle=
 \left (
 \ba
  0\\
  \vdots \\
\cdot\cdot\cdot\\
 0\\
 1\\
 0\\
 \vdots
 \ea
 \right )
 \ee
etc.

\section{Relativistic representations of
 $sl(1|1)$
 \label{valgebra} }

\subsection{
 Pseudo-Hermitian
 $\hat{h}_{(L,R)}^{(PB)}$
 and their refined partitioning\label{ponadKhare}}

In the spirit of our introductory considerations and in a way
prepared by the preceding text, let us now replace the $sl(1|1)$
commutation/anticommutation rules (\ref{antiko}) by their less
symmetric version
 \be
 \{Q,Q\}=
 \{\tilde{Q},\tilde{Q} \}=0,  \ \ \ \
 \left [ H ,Q \right ] = \left [ H, \tilde{Q} \right ] =0\,
 ,\ \ \ \ \
 \{Q,\tilde{Q}\}= H \neq H^\dagger\,.
 \label{chemiko}
 \ee
The validity of this multiplication table may be guaranteed via a
slight modification of eq. (\ref{rovni}),
 \be
 Q=\left (
 \begin{array}{cc}
 0&0\\
 \hat{a}&0
 \ea
 \right ), \ \ \ \
 \tilde{Q}=\left (
 \begin{array}{cc}
 0&\hat{c}\\
 0&0
 \ea
 \right ), \ \ \ \
 H=\left (
 \begin{array}{cc}
 \hat{h}_L&0\\
 0&\hat{h}_R
 \ea
 \right )\,.
 \label{nerovni}
 \ee
The weakening of the Hermiticity assumption means that we must
re-analyze the SUSY-induced factorization conditions in the
Hilbert space ${\cal S}={\cal R}_{(L)} \oplus {\cal R}_{(R)} $,
 \be
  H=\left (
 \begin{array}{cc}
 \hat{h}_L&0\\
 0&\hat{h}_R
 \ea
 \right )=\left (
 \begin{array}{cc}
 \hat{c}\hat{a}&0\\
 0&\hat{a}\hat{c}
 \ea
 \right )
 \,.
 \label{surovni}
 \ee
We shall try to satisfy it via a refined partitioning of both the
supercharges and of the related non-Hermitian sub-Hamiltonians
which act in the two equivalent subspaces ${\cal R}={\cal
R}_{(L,R)}$ of ${\cal S} $,
 \be
 \hat{a}=\left (
 \begin{array}{cc}
 \alpha&0\\
 0&\beta
 \ea
 \right ), \ \ \ \ \ \ \
 \hat{c}=\left (
 \begin{array}{cc}
 0&\gamma \\
 \delta&0
 \ea
 \right )\,
 \label{pofdrovni}
 \ee
 \be
  \hat{h}_{R}=\hat{c}\hat{a}=\left (
 \begin{array}{cc}
 0&\gamma\beta\\
 \delta\alpha&0
 \ea
 \right ), \ \ \ \
  \hat{h}_{L}=\hat{a}\hat{c}=\left (
 \begin{array}{cc}
 0&\alpha\gamma \\
 \beta\delta&0
 \ea
 \right )
 \,.
 \label{porovni}
 \ee
In addition to this algebra, one should return to the differential
operators, imagining that a new physical meaning must also be
assigned to all the new symbols. In the rest of this paper we
shall insert the relativistic, Klein-Gordon differential form of
the sub-Hamiltonians and show that and how all this scheme works
with the right and regular $ \hat{h}_{(R)}=\hat{h}_{(R)}^{(PB)}$
and with its left (and, as we shall see, irregular and
appropriately regularized) partner $
\hat{h}_{(L)}=[\hat{h}_{(L)}^{(PB)}]_{(reg.)}$.

\subsection{Factorizations of the regular
 $\hat{h}^{(PB)}_{(R)}$
 \label{cteniapsani}}

\subsubsection{Algebraic factorizations in eigen-bases}

Before moving to the differential-operator constructions, let us
once more switch to the linear-algebraic and matrix notation,
first employed in subsection \ref{exampleh} and then described in
full detail in Appendix A. As long as we demand, first of all,
that $ \hat{h}^{(PB)}_{(R)}=\hat{a}\hat{c}$ we have to deal, due
to eq. (\ref{porovni}), with the following two requirements,
 \be
 \hat{H}^{(KG)}_{(R)}=\alpha\cdot \gamma> 0\,, \ \ \ \ \ \ \
 \beta\cdot \delta=I\,.
 \label{identities}
 \ee
Recollecting our experience with the non-relativistic cases we may
now try to reproduce $\hat{H}^{(KG)}_{(R)}$, in the basis of its
own eigenvectors, as a diagonal matrix with elements
$\varepsilon_{(R)n}$. Most easily, this will follow from the most
elementary choice of the elements
 \be
 a_n=c_n=\sqrt{
  \varepsilon_{(R)n}  }\,,
  \ \ \ \ \ \ \
   \ \ n = 0, 1, \ldots\,
  \ee
in
 \be
 \alpha =
 \left (
 \begin{array}{ccccc}
 0&a_0&0&\ldots&\\
 0&0&a_1&0&\ldots\\
 \vdots&\ddots&\ddots&\ddots&\ddots
 \ea
 \right ), \ \ \ \ \ \ \
 \gamma =
 \left (
 \begin{array}{cccc}
 0&\ldots&&\\
 c_0&0&\ldots&\\
 0&c_1&0&\ldots\\
 \vdots&\ddots&\ddots&\ddots
 \ea
 \right )\,.
 \ee
The second identity in eq. (\ref{identities}) will then be
satisfied when we postulate, say,
 \be
 \delta =
 \left (
 \begin{array}{cccc}
 0&\ldots&&\\
 d_0&0&\ldots&\\
 0&d_1&0&\ldots\\
 \vdots&\ddots&\ddots&\ddots
 \ea
 \right )\,\ \ \ \ \ \ \ \
 \beta =
 \left (
 \begin{array}{ccccc}
 0&1/d_0&0&\ldots&\\
 0&0&1/d_1&0&\ldots\\
 \vdots&\ddots&\ddots&\ddots&\ddots
 \ea
 \right )\,
 \ee
These rules must remain compatible with the rest of eq.
(\ref{porovni}) of course.

\subsubsection{The differential-operator factorizations
 \label{hopsani}}

Besides the harmonic-oscillator illustration of the
non-relativistic SUSY QM  (cf. subsection \ref{exampleh}) we could
have used the so called Scarf example which is also described
thoroughly in the review \cite{CKS} and which may be derived from
a slightly more complicated superpotential
 \be
 W(x)=\frac{A\,\sinh \omega x +B}{\cosh \,\omega x}\,.
 \ee
We preferred to recall a slightly modified version of this
alternative solvable example sooner, viz., in the role of an
explicit sample of the Klein-Gordon operator in subsection
\ref{examplesc}. The reason was that the Scarf potential may be
treated as a bounded and arbitrarily small perturbation of the
current free Klein-Gordon field with no interaction at all. Now we
might add a comment that after a transfer of the Scarf's model to
the non-relativistic SUSY QM, the ``right" partner potential in
the general recipe (\ref{operators}) is obtained simply by the
replacement of the ``left" parameter $A$ [in the explicit formula
(\ref{11}) as well as in the subsequent solutions] by $A-\omega$.
This property is called shape invariance and its thorough
discussion may be found in ref. \cite{CKS}.

On this background one can quickly deduce the most important part
of the explicit {\em differential} version of the {\em algebraic}
factorization rules (\ref{identities}). It is given by the
innovated, {\em relativistic} formulae
 \be
 \alpha=\beta=\p_x+W(x), \ \ \ \ \ \
 \gamma=-\p_x+W(x), \ \ \ \ \ \
 \hat{H}_{L/R}^{(KG)}=
 -\p^2_x+W^2(x) \mp W'(x)\,.
 \label{operatorsKG}
 \ee
The concept of shape invariance is transferred to the relativistic
context without any perceivable change. The only {\em real}
difference concerns the Green's-function operator $\delta$ which
is defined by the second identity in (\ref{identities}). It may
have a more complicated integral-operator form but in the
factorization of the operator $\hat{h}^{(PB)}_{(R)}$ its role (of
a right pseudo-inverse of $\beta$) is purely formal.

\subsection{The final regularization and factorization of
$\hat{h}^{(PB)}_{(L)}$ }

The possible presence of an exceptional point
$\varepsilon_{(L)0}=0$ in the spectrum of the operators
$\hat{h}^{(PB)}_{(L)}$ was detected in subsection \ref{examplesc}.
As a consequence we have to be prepared to modify the whole
picture and to employ, if necessary, a transition to the
regularized Hamiltonians $[\hat{h}^{(PB)}_{(L)}]_{(reg.)}$ which
would possess a more consistent physical interpretation. Keeping
this in mind, we may recollect subsection \ref{ponadKhare} and
expect that the ``left" operator $\hat{h}^{(PB)}_{(L)}$ or rather
$[\hat{h}^{(PB)}_{(L)}]_{reg.}= \hat{c}\hat{a}$ acting in ${\cal
R}_{(L)}$ will be {\em the} desired SUSY partner of the above
``right" and regular operator $\hat{h}^{(PB)}_{(R)}=
\hat{a}\hat{c}$ defined in ${\cal R}_{(R)}$. The respective
spectra $\{\varepsilon_{(L,R)n}\}$ of these two operators must be
related by the isospectrality relation (\ref{isos}),
 \be
 \varepsilon_{(L)(n+1)}=
 \varepsilon_{(R)n}, \ \ \ \ \ \ \ \
 n = 0, 1, \ldots, \ \ \ \ \ \ \ \
 \varepsilon_{(L)0}=0\,.
 \ee
The former partner operator is defined by the product formula
 \be
 \left [
 \hat{h}^{(PB)}_{(L)}
 \right ]_{(reg.)}= \hat{c}\hat{a}=
 \left (
 \begin{array}{cc}
 0&\gamma\beta\\
 \delta\alpha&0
 \ea
 \right )\,
 \label{35}
 \ee
in the Hilbert subspace ${\cal R}_{(L)}={\cal H}_{(U)}\oplus {\cal
H}_{(D)} $ of the whole space ${\cal S}={\cal R}_{(L)}\oplus {\cal
R}_{(R)} $ where the supersymmetry is to be represented. This
means that we have to complement the above relations
(\ref{identities}) by the last two missing factorization rules in
the ``smallest" Hilbert spaces ${\cal H}$. The first one gives the
infinite-dimensional matrix formula for the product $\gamma\cdot
\beta=\hat{H}^{(KG)}_{(L)}$ in its own eigenbasis,
 \be
 \left (
 \begin{array}{cccc}
 0&\ldots&&\\
 c_0&0&\ldots&\\
 0&c_1&0&\ldots\\
 \vdots&\ddots&\ddots&\ddots
 \ea
 \right ) \cdot
 \left (
 \begin{array}{cccc}
 0&1/d_0&0&\ldots\\
 0&0&1/d_1&\ldots\\
 0&0&0&\ddots \\
 \vdots&\ddots&&\ddots
 \ea
 \right )=\left (
 \begin{array}{cccc}
 \varepsilon_{(L)0}&0&\ldots&\\
 0&\varepsilon_{(L)1}&0&\ldots\\
 0&0&\varepsilon_{(L)2}&\ddots\\
 \vdots&\ddots&\ddots&\ddots
 \ea
 \right ) \,.
 \label{newidentities}
 \ee
We see that we have to fix $\varepsilon_{(L)0}=0$ while the
consistency of our considerations follows from the resulting
$\varepsilon_{(L)n}=c_{n-1}/d_{n-1}=\varepsilon_{(R)(n-1)}$ at
$n=1,2,\ldots$. Once we choose $d_n\cdot a_n=1$, the second matrix
product evaluates to the projector
 \be
 \delta\cdot \alpha=\left (
 \begin{array}{cccc}
 0&\ldots&&\\
 d_0&0&\ldots&\\
 0&d_1&0&\ldots\\
 \vdots&\ddots&\ddots&\ddots
 \ea
 \right )
  \cdot
 \left (
 \begin{array}{ccccc}
 0&a_0&0&\ldots&\\
 0&0&a_1&0&\ldots\\
 0&0&0&a_2&\ddots \\
 \vdots&\ddots&&\ddots&\ddots
 \ea
 \right )=\left (
 \begin{array}{cccc}
 0&0&\ldots&\\
 0&1&0&\ldots\\
 0&0&1&\ddots\\
 \vdots&\ddots&\ddots&\ddots
 \ea
 \right )
 \,
 \ee
expressible, in the notation of Appendix A, as a perturbed unit
operator,
 \be
 \Pi=\left (
 \begin{array}{cccc}
 0&0&\ldots&\\
 0&1&0&\ldots\\
 0&0&1&\ddots\\
 \vdots&\ddots&\ddots&\ddots
 \ea
 \right )
 =I-|D_0\}\,\frac{1}{\{ \{ L_0|D_0\}}\,\{ \{ L_0|
  \,.
 \label{projek}
  \ee
We might conclude that a full {\em mathematical} analogy with the
non-relativistic SUSY QM of ref. \cite{CKS} is being achieved at
the cost of the regularization $I \to \Pi$ of our Hamiltonian's
action in ${\cal R}_{(L)}$. On the same purely mathematical level,
the limiting transition $\varepsilon_{(L)0} \to 0$ has a natural
consequence $|U_0\} \to 0$ and, hence, $\varphi_1(x,t) \to 0$.
This might have been expected in advance because the related
wavefunction $\varphi_2(x,t)$ cannot depend on time at
$E=\sqrt{\varepsilon_0}=0$. For this reason, the original
solutions $\varphi_{1,2}^{(PB)}(x)$ remain unchanged when the unit
operator $I$ in the relation (\ref{jepiodnadfr}) is replaced by
the projector $\Pi$.

{\it Vice versa}, {\em before} the regularization $I \to \Pi$ in
the operator $\hat{h}^{(PB)}_{(L)}$, both the sets of its
respective right and left eigenvectors  $ |n^{(\pm)} \rangle $ and
$ \langle \langle n^{(\pm)}| $ were not complete because the two
$n=0$ kets $ |0^{(\pm)} \rangle $ (as well as bras $ \langle
\langle 0^{(\pm)}|$) ceased to be independent in the limit
$\varepsilon_0 \to 0$. {\it After} the regularization $I \to \Pi$
there emerge the {\em } new $n=0$ independent pairs of
eigenvectors
 \be
 |0^{(\pm)} \rangle =
 \left (
 \ba
 \pm  \sqrt{\eta}\,|D_0\}\\ \, |D_0\}
 \ea
 \right ), \ \ \ \ \ \
 \langle \langle 0^{(\pm)} |
 = \left ( \{ \{ L_0|
 , \pm  \sqrt{\eta}\, \{ \{ L_0|
 \right ) \,
 \label{generatedbe}
 \ee
(with any $\eta>0$)  which make the sets of eigenvectors complete
in ${\cal R}$. Then we may freely apply all the formulae provided
by Appendix A in the regular case. The new ``missing" vectors
(\ref{generatedbe}) do not play any important role after all. They
do not even enter the spectral representation of the block
$\hat{H}^{(PB)}_{(L)}$ or projector $\Pi$ in our factorized
Klein-Gordon SUSY partner $[\hat{h}^{(PB)}_{(L)}]_{(reg.)}$ of
$\hat{h}^{(PB)}_{(R)}$ at all.

\section{Summary  \label{fiveth}}

Our present extension of the formalism of SUSY QM to the
relativistic domain is methodically promising and conceptually
transparent. It preserves a close similarity between Klein Gordon
eq. (\ref{jdvar}) and the usual Schr\"{o}dinger equation
(\ref{SE}). We found a technical key to their parallelism in a
refined partitioning of the Hilbert space in the relativistic
case. We also avoided the main obstacle of the relativistic  SUSY
QM construction which lied in the difficult {\em physical}
interpretation of the vanishing ground-state or exceptional-point
eigenvalue $\varepsilon_{(L)0}=0$.

In the context of the recent studies of the Klein-Gordon equation,
the emergence of the EP difficulty is new. The reason is that a
strong attraction must be present in our $\hat{H}^{(KG)}_{(L)}$ in
order to produce the eigenvalue which lies so deeply below the
boundary of the continuous spectrum $m^2(\infty)=m_0^2$. The
exceptional vanishing value of $\varepsilon_{(L)0}$ corresponds to
the strength of interaction where a small additional perturbation
is already able to produce a collapse or, in the language of the
relativistic quantum field theory, a spontaneous pair-creation,
etc.

One of the key merits of non-relativistic SUSY QM is the
explanation of the origin of the {exact solvability} of many
traditional Schr\"{o}dinger equations. In the future, our SUSY
construction might play a similar mathematical role in the
relativistic context. Its key features parallel the list of
obstacles in our introductory section:

\begin{itemize}

\item (a) The range of our Klein-Gordon SUSY models is constrained
to the ``no-collapse" regime where the energies remain real. We
cannot describe the effects like a spontaneous creation of
particle-antiparticle pairs of course. The natural boundary of our
formalism is clearly marked by the spontaneous complexification of
the energy pairs.

\item (b) Phenomenological aspects of our present SUSY formalism
are transparent and similar to their non-relativistic
predecessors. A notable exception concerns the unclear
interpretation of the vanishing energies, circumvented here by a
regularization of the ``left" sub-Hamiltonians $\hat{h}^{(PB)}_L$.

\item (c) Mathematical feasibility of the factorization of our
present relativistic Klein-Gordon Hamiltonians has been shown
rendered possible by the {\em same} formal regularization as
above. This coincidence may be understood as a quantum mechanical
analogue of consistency of the high-energy cut-off in quantum
field theory.

\end{itemize}

 \noindent
Marginally, let us point out the possible methodical relevance of
the present regularization of the level crossing at
$\varepsilon_0=0$ for an improvement of our future understanding
of some models in non-quantum domains of physics where the very
similar ``exceptional-point" singularities have been observed
inside the intervals of variability of relevant parameters
\cite{Heiss,Uwe}.

\section*{Acknowledgement}

Work partially supported by the grant Nr. A 1048302 of GA AS CR.


\newpage

\section*{Appendix A: Biorthogonality
and completeness
 \label{ctyri}\label{geometrie}}

In the notation of section \ref{dveje} (omitting, temporarily, the
``pathologically" vanishing $\varepsilon_0=0$), the results of the
solution of the two equations
 \be
 \hat{h}^{(PB)}\,|n^{(\pm)} \rangle =
 E_n^{(\pm)}\,|n^{(\pm)} \rangle , \ \ \ \ \ \ \ \
 \langle \langle n^{(\pm)} |\,
 \hat{h}^{(PB)} =
 E_n^{(\pm)}\,
 \langle \langle
 n^{(\pm)} | \,
 \label{bigger}
 \ee
may be well assumed to span the Hilbert space ${\cal R}$. As long
as the superscripts $^{(\pm)}$ may enter just the norms, we may
abbreviate $| D_n^{(\pm)}\}= \nu_{n}^{(\pm)}\,| D_n\}$ and  $|
L_n^{(\pm)}\} \}= \kappa_n^{(\pm)}\,| L_n\} \}$ and treat the two
reduced equations
 \be
 \hat{H}^{(KG)}\,|
 D_n\}=\varepsilon_n\, |D_n\},
  \ \ \ \ \ \  \ \ \ \ \{ \{ L_n|\,\hat{H}^{(KG)}
 =\varepsilon_n\, \{ \{ L_n|
 , \ \ \ \ \ \ n = 0, 1, \ldots \,
 \label{smaller}
 \ee
as the source of the two complete sets in ${\cal H}$.

\subsection*{A. 1. Smaller Hilbert spaces
\label{yudve}${\cal H}$}

Functions defined by the two eqs. (\ref{smaller}) may be used to
pre-multiply the partner equation. The subsequent subtraction of
the resulting matrix elements may be easily checked to give the
rule
 \be
 0 = \left (\varepsilon_m-\varepsilon_n\right )\,
 \{ \{ L_m | D_n\}\,.
 \ee
This is to be read as a bi-orthogonality relation in the regular
space ${\cal H}_{(R)}$ where $\varepsilon_0>0$. The overlaps $\{
\{ L_m | D_n\}$ must necessarily vanish whenever the energies
remain non-degenerate or at any pair of subscripts such that
$\varepsilon_m \neq \varepsilon_n$. For the sake of a
simplification of our present notation we shall skip the
degenerate case as not sufficiently interesting and abbreviate $
\varrho_{n}=\{ \{ L_n| D_n\}$. Assuming also that these overlaps
do not vanish we may write the unit operator in ${\cal H}_{(R)}$
in the form of the decomposition
 \be
 I = \sum_{n=0}^\infty\,|D_n\}\,\frac{1}{\varrho_n}\,\{ \{ L_n|\,.
 \ee
The replacement of this operator by the projector $\Pi$ of eq.
(\ref{projek}) proves trivial,
 \be
 \Pi = \sum_{n=1}^\infty\,|D_n\}\,\frac{1}{\varrho_n}\,\{ \{ L_n|\,.
 \label{nevyjek}
 \ee
Finally, the spectral decomposition of the Hamiltonian remains
also very natural and transparent,
 \be
 \hat{H}^{(KG)}
  = \sum_{n=0}^\infty\,|D_n\}\,\frac{\varepsilon_n}{\varrho_n}\,\{
  \{ L_n|\,.
 \ee
The second item in eq. (\ref{smaller}) may be also written in its
Hermitian conjugate form, i.e., as the relation
 \be
 \hat{H}^{(KG)}\,{G}^{-1}
 |L_n \} \}=\varepsilon\, {G}^{-1}\,|L_n \} \},
  \ \ \ \ \ \ \varepsilon =E^2>0
 \,.
 \label{upedhhdfr}
 \ee
Its comparison with the first item in the same equation reveals
that only a multiplicative constant may distinguish between these
two alternative forms of the eigenkets of the same operator,
 \be
 |L_n \} \} = q_n\,{G}\,|D_n \}\,.
 \ee
For the sake of simplicity, let us require that  $q_n$ (which
might be called generalized quasi-parity) is real. Then our latter
formula defines the left-action eigenvectors in terms of the
right-action ones (or vice versa) in the halved Hilbert space.
Now, it is important to imagine that our choice of the value of
the quasi-parity $q_n$ determines fully the above-introduced
overlaps
 \be
 \varrho_n=\{ \{ L_n|D_n\} \equiv q_n\,\{ D_n| {G} | D_n\}\,,
 \ee
(and, in particular, their signs) because the values of the matrix
elements $\{ D_n| {G} | D_n\}$ themselves are real (the metric
${G}$ is always Hermitian) and may be considered known (or
calculated, by integration) in advance.

\subsection*{A. 2. Larger Hilbert
spaces \label{ygudve} ${\cal R}$}

The two-component functions defined by eqs. (\ref{bigger}) in the
Hilbert spaces ${\cal R}$ may be used again as pre-multiplication
factors which convert the subtracted matrix elements into another
fundamental biorthogonality condition
 \be
 0 = \left (\tau\,\sqrt{\varepsilon_m}-
 \sigma\,\sqrt{\varepsilon_n} \right )\,
 \langle \langle m^{(\tau)} | n^{(\sigma)}\rangle\,.
 \ee
As long as ${\cal R}={\cal H}\oplus {\cal H}$, we could just
insert and use the results of the preceding subsection.
Alternatively, our notation enables us to shorten such a procedure
and deduce that the overlaps $\langle \langle m^{(\tau)} |
n^{(\sigma)}\rangle$ must necessarily vanish whenever the new
energies $E_n^{(\pm)}$ remain non-degenerate. The latter condition
means that the overlaps are zero not only when $\varepsilon_m \neq
\varepsilon_n$ but also when the signatures do not coincide, i.e.,
whenever $\pm 1 =\tau  \neq \sigma = \mp 1$.

In a way paralleling our previous text we abbreviate now $
\mu_n^{(\tau)}=\langle \langle n^{(\tau)} | n^{(\tau)}\rangle$ and
expand the unit operator in the larger space ${\cal R}$,
 \be
 I = \sum_{n=0}^\infty\,\sum_{\tau = \pm 1}| n^{(\tau)}\rangle
 \,\frac{1}{\mu_n^{(\tau)}}\, \langle \langle n^{(\tau)} | \,.
 \ee
With $E_n^{(\tau)}=\tau\,\sqrt{\varepsilon_n}$, the parallel
spectral decomposition of the Hamiltonian reads
 \be
 \hat{h}^{(PB)}
  = \sum_{n=0}^\infty\,\sum_{\tau = \pm 1}| n^{(\tau)}\rangle
 \,\frac{E_n^{(\tau)}}{\mu_n^{(\tau)}}\, \langle \langle n^{(\tau)} | \,.
 \ee
The use of the pseudo-Hermiticity rules (\ref{etaf}) +
(\ref{neheri}) transforms again the equation for the Hermitian
conjugate double-bra vectors into the following equivalent
equation for the double-kets,
 \be
 \hat{h}^{(PB)}\,{\cal G}^{-1}
 | n^{(\tau)}\rangle\rangle=E\,
 {\cal G}^{-1}\,| n^{(\tau)}\rangle\rangle,
  \ \ \ \ \ \ E = \tau\,\sqrt{\varepsilon}
 \,.
 \label{upjjhdfr}
 \ee
This means that another multiplicative real constant
$Q_n^{(\tau)}$ is to be introduced in order to distinguish between
the eigenfunctions,
 \be
 | n^{(\tau)}\rangle\rangle = Q_n^{(\tau)}
 \,{\cal G}\,| n^{(\tau)}\rangle\,.
 \ee
This formula again implies that
 \be
 \mu_n^{(\tau)}=\langle \langle n^{(\tau)} |n^{(\tau)}\rangle
  \equiv Q_n^{(\tau)}\,\langle n^{(\tau)} | {\cal G}
   |n^{(\tau)}\rangle\,.
 \ee
After we recall all the definitions, we may already evaluate
 \be
 \mu_n^{(\tau)}=2\,\tau\,\sqrt{\varepsilon_n} \varrho_n\,.
 \ee
In the conclusion let us return to our initial assumptions and try
to admit that the ground-state-like eigenvalue $\varepsilon_0= 0$
vanishes. Then, equation (\ref{generated}) would merely define the
two identical vectors at both signs $^{(\pm)}$ at $n=0$. Such a
feature is characteristic for the non-diagonalizable (usually
called Jordan-block) limits of non-Hermitian operators
\cite{Iochvidov}. In the spectra of differential operators these
points are also known as ``Bender-Wu singularities" \cite{BW}, as
the points of an ``unavoided level-crossing" \cite{ptho} or simply
as ``exceptional points" \cite{Heiss}. In our present paper, their
properties will only be derived via a limiting transition
$\varepsilon_0\to 0^+$ from the regular domain.

\end{document}